%% file: charm.tex
\documentclass[12pt]{article}
\usepackage{graphicx}
\include{babarsym}

\usepackage{multirow}
\usepackage{array}

\RequirePackage{xspace}
\usepackage{relsize}

\def\pbnr{}
\def\speaker{Nicola Neri}
\def\onbehalfof{the \babar\ Collaboration}
\def\title{Searches for \CP violation in charm decays at \babar}
\def\affiliation{INFN, Sezione di Milano, Milano, Italy\\
}

\def\beq{\begin{equation}}
\def\eeq{\end{equation}}
\def\eeqn{\end{equation}}

\def\beqa{\begin{eqnarray}}
\def\eeqan{\end{eqnarray}}

\def\babar {\mbox{\slshape B\kern-0.1em{\smaller A}\kern-0.1em
    B\kern-0.1em{\smaller A\kern-0.2em R}}}
\def\CP{\ensuremath{C\!P}\xspace}
\def\CPV{\ensuremath{C\!PV}\xspace}
\def\Dkkpi    {\ensuremath{D^{+}\to K^{+}K^{-}\pi^{+}}\xspace}
\def\AGamma{\ensuremath{A_\Gamma}\xspace}

\input charmmacros.tex

\begin{document}
\begin{titlepage}
\pubblock

\vfill
\Title{\title}
\vfill
\Author{\speaker
\OnBehalf{\onbehalfof}}
\Address{\affiliation}
\vfill
\begin{Abstract}

\end{Abstract}
\begin{abstract}
In the Standard Model \CP violation in charm decays is expected to be very small, at the level of 0.1\% or less.
A significant excess of \CP violation with respect to the Standard Model predictions would be a signature of new physics.  
We report on recent searches for \CP violation in charm meson decays at \babar, using a data sample 
 corresponding to an integrated luminosity of about 470 \invfb. 
In particular, we report on searches for \CPV in the 
3-body $\Dp\to\Kp\Km\pip$ decay
and for decay modes with a \KS 
in the final state, such as  $\Dp\to\KS\Kp$, $\Ds\to\KS\Kp$, $\Ds\to\KS\pip$. 
A lifetime ratio analysis of $\Dz\to\Kp\Km, \pip\pim$ with respect to $\Dz\to\Km\pip$ 
decays, which is sensitive to \Dz-\Dzb mixing and \CP violation, is also presented here.
\end{abstract} 

\vfill
\begin{Presented}
\venue
\end{Presented}
\vfill
\end{titlepage}
\def\thefootnote{\fnsymbol{footnote}}
\setcounter{footnote}{0}
%


\section{Introduction}
\label{sec:intro}
In the Standard Model (SM) \CP violation (\CPV) is accommodated by the CKM~\cite{Cabibbo:1963yz,Kobayashi:1973fv} 
mechanism which regulates the mixing of the three families of quarks. 
This mechanism has been proved to work well according to the experimental results 
 that have been provided mostly by the \B-Factory experiments, \babar\ and Belle, during the last decade. However, the CKM mechanism 
 is not sufficient to describe the absence of antimatter in the universe, and so this represents an open question for both 
 experimental and theoretical physicists. Other sources of \CPV are currently under investigation, and the 
 charm sector represents an interesting probe for this purpose.
\par
In charm meson decays \CPV is expected to be at the level of 0.1\% or less~\cite{Bianco:2003vb,Grossman:2006jg}, 
 although the predictions are affected by large theoretical uncertainties due to long distance interactions. 
%
The study of \CPV in singly Cabibbo-suppressed (SCS) charm decays is particularly sensitive to 
new physics (NP)~\cite{Grossman:2006jg}, while evidence of indirect \CPV 
in \Dz-\Dzb mixing, with the current experimental precision, would be a clear sign for NP. 
Throughout the following discussion the use of charge conjugate reactions is implied, unless otherwise indicated.
\section{Search for direct \CPV in \boldmath{$D^+ \to K^+K^-\pi^+$} decay}
The \babar\ experiment has recently searched for \CPV in the singly Cabibbo-suppressed 
$D^+ \to K^+K^-\pi^+$ decay using a data sample corresponding to an integrated luminosity 
of 476~\invfb~\cite{Lees:2012nn}.
The 3-body decay  proceeds mainly through quasi-two-body decays with resonant intermediate states, 
which allows the investigation of the Dalitz plot substructure for asymmetry in both magnitude and phase for each  
intermediate state. In the search for \CPV,  5 different approaches were adopted: 
a measurement of the integrated \CP asymmetry,  
 a measurement of the \CP asymmetry in four regions of the Dalitz plot, 
a comparison of the binned $D^+$ and $D^-$ Dalitz plots, 
 a comparison of the Legendre-polynomial-moment weighted distributions in the 
$K^+K^-$ and $K^-\pi^+$ systems, and a comparison of 
 the results of a parameterized fit to the $D^+$ and $D^-$ Dalitz plots. 
Only the last one is model-dependent, while the previous four approaches are model-independent.
\par
The signal yield is about 223,700 events, with  signal purity of about 92\%. 
%
The \CP-violating decay rate asymmetry, $A_{\CP}$, was determined to be $(0.37 \pm 0.30\textrm{(stat)} \pm 0.15\textrm{(syst)})\%$.
The \CP asymmetries in different regions of the Dalitz plot, defined by the reconstructed invariant mass squared values 
$m^2(K^-K^+)$ and $m^2(K^-\pi^+)$, are reported in Table~\ref{tab:KKpi_DPregions}.
\begin{table}[htb]
\begin{center}
\setlength{\extrarowheight}{1.5pt}
\scalebox{0.9}{%
\begin{tabular}{|m{6.0cm}|c|}  
\hline
Dalitz plot region  & $A_{\CP}$ (\%) \\
\hline
Below $\bar{K}^{*}(892)^0$ (A)     &   $-0.7\pm 1.6\textrm{(stat)}\pm 1.7\textrm{(syst)}$ \\
$\bar{K}^{*}(892)^0$ (B)          &  $-0.3\pm 0.4\textrm{(stat)}\pm 0.2\textrm{(syst)}$ \\
$\phi(1020)$ (C)                   &  $-0.3\pm 0.3\textrm{(stat)}\pm 0.5\textrm{(syst)}$ \\
Above $\bar{K}^{*}(892)^0$ and $\phi(1020)$ (D) &  $\phantom{-}1.1\pm 0.5\textrm{(stat)}\pm 0.3\textrm{(syst)}$ \\
\hline
\end{tabular}
}
\caption{\CP asymmetry in the regions (A), (B), (C) and (D) of the Dalitz plot shown in Fig.~\ref{fig:dpfit}.
 The first error is statistical and the second is systematic.}
\label{tab:KKpi_DPregions}
\end{center}
\end{table}
Model-independent techniques were used to search for \CPV in the Dalitz plot.
These were based on 
a comparison of the binned $D^+$ and $D^-$ Dalitz plots, and on a comparison of the Legendre-polynomial-moment weighted distributions 
in the $K^+K^-$ and $K^-\pi^+$ systems. 
The distributions of normalized residuals in equally 
populated bins ($\sim1000$ events per bin) of the $D^+$ and $D^-$ Dalitz plots
were fitted with a Gaussian function. The fit yielded a mean of $0.08 \pm 0.15$ and a 
r.m.s. deviation of $1.11 \pm 0.15$, which corresponds to a probability
of 72\% that the two Dalitz plots are consistent with no \CPV. 
The comparison of Legendre-polynomial-moments for the $K^+K^-$ and $K^-\pi^+$ 
systems separately was found to be  
consistent with no \CPV with a probability of 11\% and 13\%, respectively.
\par
A model-dependent technique based on a comparison of parameterized fits to the two Dalitz plots was also used 
 to search for \CPV. The  $D^+$ decay amplitude was parameterized as a coherent sum of amplitudes describing
 the relevant two-body intermediate states (16 resonances) plus a constant amplitude over the Dalitz plot for the 
 non-resonant (NR) contribution. 
\begin{figure}[htb]
\begin{center}
\includegraphics[width=0.75\textwidth]{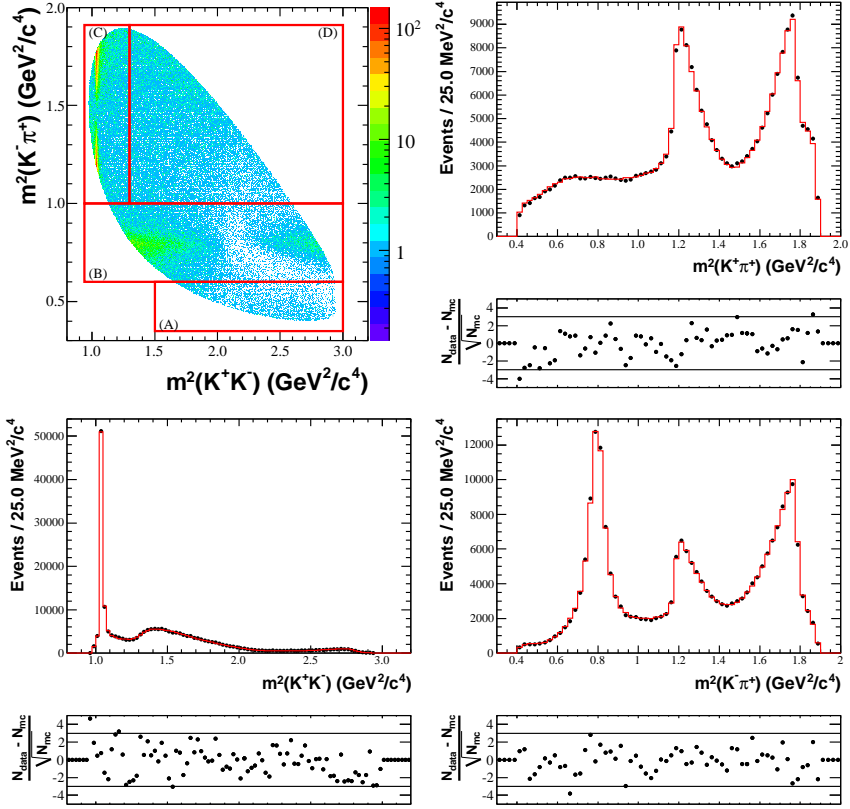}
\caption{\Dkkpi Dalitz plot and fit projections assuming no {\CP}V, with the
  regions used for model-independent comparisons also indicated as
  boxes. For each projection, the data are represented by points with errors, and the fit result
  by the histogram. The normalized residuals below each histogram, defined as
  $(N_{Data} - N_{MC})/\sqrt{N_{MC}}$, lie between $\pm5\sigma$.}
\label{fig:dpfit}
\end{center}
\end{figure}
%
The resonances that contribute to the fit with the largest fit fractions are  
 the $\phi(1020)$ $(28.42\pm0.13)\%$, $\bar{K}^{*}(1430)^0$ $(25.32\pm2.24)\%$, and 
the  $\bar{K}^{*}(892)^0$ $(21.15\pm0.20)\%$.
The results of the fit to the $D^+$ and $D^-$ Dalitz plots do not show evidence 
of \CPV for the following amplitudes:  
$\bar{K}^{*}(892)^0 \Kp$, $\bar{K}^{*}(1430)^0 \Kp$,
 $\phi(1020) \pip$, NR, $\bar{\kappa} (800)^0 \Kp$, $a_0(1450)^0 \pip$, $f_0(980) \pip$, $f_0(1370) \pip$. 
%
%
%
\section{Search for \CPV in \boldmath{$D^+ \to \KS K^+$} and \boldmath{$D_s^+ \to \KS K^+, \KS \pi^+$} decays}
\label{sec:CPV_DtoKsh}
In $D$ meson decays with a \KS in the final state, \CP-violating asymmetries defined as 
\beq
A_{\CP} = \frac{\Gamma(D^+_{(s)}\to \KS h^+) - \Gamma(D^-_{(s)}\to \KS h^-)}
{\Gamma(D^+_{(s)}\to \KS h^+) + \Gamma(D^-_{(s)}\to \KS h^-)}
= A^{\Delta C}_{\CP} + A_{\CP}^{\KS}, 
\eeqn
can receive contributions from \CPV in $\Delta C =1$ quark transitions ($A^{\Delta C}_{\CP}$), 
 and from \CPV in \Kz-\Kzb mixing ($A_{\CP}^{\KS}$).
The value of the contribution from \Kz-\Kzb mixing is precisely determined to be 
$A_{\CP}^{\KS} = [\pm0.332\pm0.006]\%$~\cite{PDG:2012}, 
where the $\pm$ sign depends on whether a \Kz (+) or a \Kzb (-) 
is produced in the decay.
The SM prediction has to be corrected for the detector acceptance as a function of the decay time~\cite{Grossman:2011zk}, and
the correction is at the level of few percent at the $B$ factories.
A sizable deviation of the measured $A_{\CP}$ value from the $A_{\CP}^{\KS}$ predicted value would indicate \CPV 
in the $\Delta C =1$ quark transition, and might indicate NP effects.  
\par
 The \babar\ experiment has recently searched for 
\CP asymmetries in the  $D^+_{(s)} \to \KS K^+$ and $D_s^+ \to \KS \pi^+$ decay modes~\cite{Lees:2012jv}
using a data sample corresponding to an integrated luminosity of 469~\invfb.
%
The reconstructed asymmetry is defined as 
\beq
A_{rec} = \frac{N_{D^+_{(s)}} - N_{D^-_{(s)}}}{N_{D^{+}_{(s)}} + N_{D^-_{(s)}}}
=  A_{\CP} + A_{FB} + A_{\epsilon},  
\eeqn
where $N_{D^+_{(s)}}$ ($N_{D^-_{(s)}}$) is the number of $D^+_{(s)}$ ($D^-_{(s)}$) decays determined from 
 the fit to the relevant invariant mass distribution, 
 $A_{FB}$ is the forward-backward ($FB$) asymmetry, and $A_{\epsilon}$ is the detector-induced 
 charge reconstruction asymmetry; $A_{FB}$ originates from the $FB$ asymmetry in $\epem \to \ccbar$ 
 production, coupled with the asymmetric acceptance of the detector, and is measured 
 directly on data together with $A_{\CP}$~\cite{delAmoSanchez:2011zza}. 
The fits to the $m(\KS h)$ distributions yield $(159.4\pm 0.8)\times 10^3$ signal
events for $D^+\to\KS K^+$, $(288.2 \pm 1.1) \times 10^3$ for
$D_s^+\to\KS K^+$, and $(14.33\pm0.31)\times10^3$ for $D_s^+\to\KS \pi^+$.
\par
The \CP-violating asymmetries $A_{\CP}$ for the 
$D^+ \to \KS K^+$, $D_s^+ \to \KS K^+$, and $D_s^+ \to \KS \pi^+$ decays
are determined to be 
$[0.13\pm 0.36\textrm{(stat)}\pm 0.25\textrm{(syst)}]\%$, 
$[-0.05\pm 0.23\textrm{(stat)}\pm 0.24\textrm{(syst)}]\%$, and
$[0.6\pm 2.0\textrm{(stat)}\pm 0.3\textrm{(syst)}]\%$, respectively. 
The primary source of systematic error is due to the statistical uncertainty in the  
 determination of the charge asymmetry in track reconstruction efficiency.
\par
The contribution to the \CP asymmetries due to 
 the $\Delta C =1$ transition is measured to be 
$[0.46\pm 0.36\textrm{(stat)}\pm 0.25\textrm{(syst)}]\%$, 
$[0.28\pm 0.23\textrm{(stat)}\pm 0.24\textrm{(syst)}]\%$, and
$[0.3\pm 2.0\textrm{(stat)}\pm 0.3\textrm{(syst)}]\%$ for the respective decay processes. 
The results are consistent with zero, and with the SM predictions within one standard deviation.
\section{Measurement of \boldmath{\Dz-\Dzb} mixing, and search for indirect \CPV in \boldmath{$\Dz \to K^+K^-$} and \boldmath{$\Dz \to \pi^+\pi^-$} decays}
\label{sec:yCP}
The \babar\  experiment has recently updated the measurement of the mixing parameter
 $y_{\CP}$ and the \CP-violation parameter $\Delta Y$~\cite{Lees:2012qh}.
The definitions of \deltaY 
\footnote{Note that this definition for \deltaY uses a different sign convention than that used in previous \babar\ publications~\cite{Aubert:2007en,Aubert:2009ai}.}
and $A_{\Gamma}$ are the following:
\beq
\deltaY = \frac{\Gamma^+ - \bar{\Gamma}^+}{2\Gamma} = (1+\yCP) \AGamma, \quad \quad \quad \quad \quad
\AGamma = \frac{\bar{\tau}^+ - \tau^+}{\bar{\tau}^+ + \tau^+},  
\eeqn
where $\tau^+=1/\Gamma^+$ $(\bar{\tau}^+=1/\bar{\Gamma}^+)$ are the effective lifetimes for \Dz (\Dzb) 
decaying to the \CP-even final states $K^+K^-$ and $\pi^+\pi^-$. 
In this analysis \CP conservation in the decay is assumed, 
and results are averaged over the $K^+K^-$ and $\pi^+\pi^-$ modes.
\par
The measurements are based on the ratio of lifetimes extracted simultaneously from a sample 
of \Dz mesons produced through the flavor-tagging process \break $D^{*+} \to \Dz \pi^+$, 
where the \Dz decays to $K^-\pi^+$, $K^-K^+$, or $\pi^-\pi^+$;
additional samples of untagged decays for $\Dz \to K^-\pi^+$ and $\Dz \to K^-K^+$ are used for the measurement 
 of \yCP. The latter have about 4 times the  statistics of the corresponding flavor-tagged samples, but have 
lower purity.
%
%
The flight length is reconstructed by means of a simultaneous kinematic fit to the decay vertex and production vertex 
of the \Dz, the latter being constrained 
to  originate within the \epem collision region.
The most probable $\sigma_t$ value is about 40\% of the nominal \Dz lifetime, and only candidates with  $\sigma_t<0.5 \ps$ are retained for the fit.
The \babar\ experiment measures  $y_{\CP} = [0.72 \pm 0.18(\textrm{stat}) \pm 0.12(\textrm{syst}) ] \%$ and 
$\Delta Y = [0.09 \pm 0.26(\textrm{stat}) \pm 0.06(\textrm{syst}) ] \%$ using a data sample corresponding 
 to an integrated luminosity of 468\invfb~\cite{Lees:2012qh}.
The measurement of \yCP\ is the most precise single measurement to date.
%
%
%
\Acknowledgements
I am grateful to my \babar\ colleagues for providing excellent results and feedback for this 
 conference, and to the conference organizers for the invitation to this interesting event.

\end{document}

%% file: charmmacros.tex
\newcommand\pubnumber{\pbnr}
\newcommand\pubdate{\today}
\def\Title#1{\begin{center} {\Large #1 } \end{center}}
\def\Author#1{\begin{center}{ \sc #1} \end{center}}

\newcommand{\OnBehalf}[1]{\sbox0{#1}\ifdim\wd0=0pt
        {}
	\else
	{\\on behalf of #1}
	\fi}
\newcommand{\SupportedBy}[1]{\sbox0{#1}\ifdim\wd0=0pt
        {}
	\else
	{\footnote{#1}}
	\fi}
\def\Address#1{\begin{center}{ \it #1} \end{center}}

\newcommand\pubblock{\includegraphics[width=5cm]{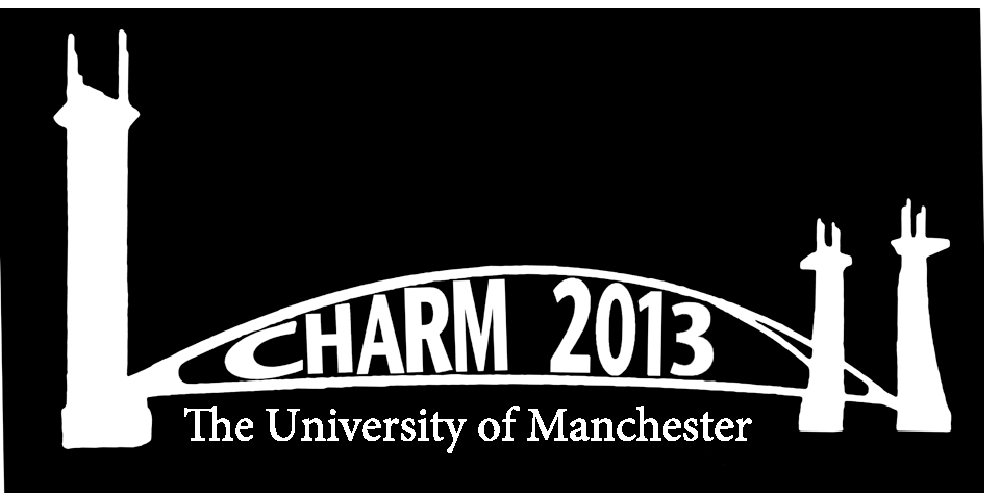}\hfill{\begin{tabular}{l} \pubnumber\\
         \pubdate  \end{tabular}}}
\newenvironment{Abstract}{\begin{quotation}  }{\end{quotation}}
\newenvironment{Presented}{\begin{quotation} \begin{center} 
             PRESENTED AT\end{center}\bigskip 
      \begin{center}\begin{large}}{\end{large}\end{center} \end{quotation}}
\def\Acknowledgements{\bigskip  \bigskip \begin{center} \begin{large}
             \bf ACKNOWLEDGEMENTS \end{large}\end{center}}
\def\venue{The 6$^{th}$ International Workshop on Charm Physics\\
(CHARM 2013)\\
Manchester, UK,  31 August -- 4 September, 2013}

\def\beq{\begin{equation}}
\def\eeq#1{\label{#1}\end{equation}}
\def\eeqn{\end{equation}}

\def\beqa{\begin{eqnarray}}
\def\eeqa#1{\label{#1}\end{eqnarray}}
\def\eeqan{\end{eqnarray}}

\let\bar=\overbar

\def\Dslash{\not{\hbox{\kern-4pt $D$}}}
\def\dslash{\not{\hbox{\kern-2pt $\del$}}}

\def\BR{\mbox{\rm BR}}

\def\msb{{\bar{\ssstyle M \kern -1pt S}}}

